\documentclass[twocolumn,reprint,amsmath,amssymb,cha,prb,aps]{revtex4-1}

\usepackage{docs}
\usepackage{epsfig}
\usepackage{graphicx}
\usepackage{bm}%
\usepackage[colorlinks=true,linkcolor=blue]{hyperref}%

\begin{document}

\title{Spectral evolution in a Shastry-Sutherland lattice, HoB$_4$}

\author{Deepnarayan Biswas$^1$}
\author{Nishaina Sahadev$^1$}
\author{Ganesh Adhikary$^1$}
\author{Geetha Balakrishnan$^2$}
\author{Kalobaran Maiti$^1$}
\altaffiliation{kbmaiti@tifr.res.in}

\affiliation{$^1$Department of Condensed Matter Physics and
Materials' Science, Tata Institute of Fundamental Research, Colaba,
Mumbai - 400 005, India.\\
$^2$Department of Physics, University of Warwick, Coventry, CV4 7AL,
UK.}

\date{\today}

\begin{abstract}
We studied the electronic structure of a Shastry-Sutherland lattice
system, HoB$_4$ employing high resolution photoemission spectroscopy
and {\em ab initio} band structure calculations. The surface and
bulk borons exhibit subtle differences, and loss of boron compared
to the stoichiometric bulk. However, the surface and bulk conduction
bands near Fermi level are found to be similar. Evolution of the
electronic structure with temperature is found to be similar to that
observed in a typical charge-disordered system. A sharp dip is
observed at the Fermi level in the low temperature spectra revealing
signature of antiferromagnetic gap. Asymmetric spectral weight
transfer with temperature manifests particle-hole asymmetry that may
be related to the exotic properties of these systems.
\end{abstract}


\maketitle

\section{Introduction}

Physics of disorder is one of the most challenging issues in
material science as it is hard to avoid disorder in real systems
arising from defects, imperfections {\em etc.} While it is hard to
probe these natural effects, some lattice structures possess
intrinsic disorder on the basis of electronic interactions leading
to exotic properties providing pathways to probe disorder induced
effects on the electronic properties. Shastry-Sutherland lattice is
one such realization, where the strong diagonal exchange
interaction, $J_2$ forms spin-dimers. Competition between $J_2$ and
inter-dimer interactions, $J_1$ leads to interesting phase diagram
involving spin-liquid phase, long-range ordered phase {\it
etc.}\cite{ssl} Plethora of exotic phases could be derived via
tuning some control parameters. E.g. J. Liu {\em et al.} showed that
charge carrier doping (electron or hole) in such systems leads to
strong asymmetry in the electronic properties - while hole doping
renormalizes spin-spin interactions leading to superconductivity,
electron doped systems can be metallic but
non-superconducting.\cite{nandini} Varieties of fields have emerged
based on these spin-disordered parent systems. Experimental
realization of these effects are not so common due to paucity of
materials exhibiting such structures. Most of the studies so far are
centered around the compound, SrCu$_2$(BO$_3$)$_2$, where Cu$^{2+}$
has spin-half state and forms the Shastry-Sutherland lattice
exhibiting interesting phenomena.\cite{miyahara}

\begin{figure}
\includegraphics [scale=0.4]{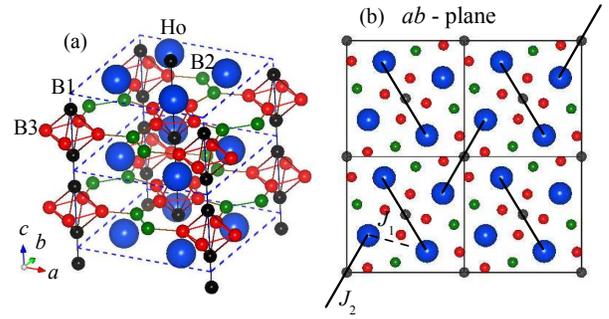}
 \vspace{-52ex}
\caption{(a) Crystal structure of HoB$_4$. (b) $ab$ plane containing
Ho and B atoms. The lines show strong antiferromagnetic coupling
between Ho sites that form spin-dimers as proposed in a
Shastry-Sutherland lattice.}
 \vspace{-4ex}
\end{figure}

Recently, it was observed that rare-earth tetraborides possess
structure and magnetic interactions akin to Shastry-Sutherland
lattice as shown in Fig. 1. The diagonal magnetic interaction, $J_2$
between the rare earth moments shown in the figure are strong
forming spin-dimers. These materials exhibit diversified electronic
and magnetic properties depending on interaction, $J_1$ among these
dimers relative to $J_2$.\cite{rb4,song} For example, CeB$_4$ and
YbB$_4$ do not exhibit long range order, while PrB$_4$ orders
ferromagnetically at low temperatures and most of the other
tetraborides order antiferromagnetically.\cite{buschow,choi} Some of
them exhibit multiple magnetic phase transitions at low
temperature.\cite{yin} For example, TbB$_4$ exhibits non-collinear
magnetic structure with two antiferromagnetic phase
transitions.\cite{matsumura} Non-collinear magnetism with quadrupole
strain and multiple phase transition is observed in
DyB$_4$.\cite{song}

HoB$_4$ is one among the complex tetraborides exhibiting multiple
magnetic transitions and non-collinear magnetism.\cite{fisk} The
crystal structure of HoB$_4$ is tetragonal\cite{yin} with $a$ =
7.085 \AA\ and $c$ = 4.004 \AA. Depending upon the symmetry of the
boron sites, there are three non-equivalent borons: B1 (apical) and
B3 (equatorial) form octahedra, and B2 forms boron dimers connecting
the boron octahedra. The interaction among Ho moments are mediated
by the borons. The spin-spin interactions in the $ab$-plane as shown
in Fig. 1(b) correspond to Shastry-Sutherland type lattice. HoB$_4$
exhibits two antiferromagnetic transitions - at T$_{N1}$ = 7.1 K, it
exhibits an incommensurate antiferromagnetic ordering. The magnetic
structure becomes commensurate at T$_{N2}$ = 5.7 K.\cite{okuyama}
Resistivity and transport measurements in the presence of magnetic
field exhibit anomalous phase diagram involving varieties of
metamagnetic phase.\cite{kim} Evidently, HoB$_4$ is complex
exhibiting plethora of interesting phases and one of the ideal
systems to study disorder induced effects in the electronic
structure. Here, we studied the electronic structure of HoB$_4$
employing high resolution photoemission spectroscopy and {\it ab
initio} band structure calculations. Interesting surface-bulk
features are observed in the experimental spectra. The evolution of
the electronic structure appears to be similar to that observed for
charge-disordered systems.

\section{Experiment}

Single crystalline sample of HoB$_4$ was grown using mirror
furnace\cite{Balakrishnan} and characterized by $x$-ray diffraction,
Laue pattern, transport and magnetic measurements revealing high
quality of the sample. Photoemission measurements were carried out
using R4000 WAL electron analyzer from Gammadata Scienta and
monochromatic photon sources, Al K$\alpha$ (1486.6 eV), He
{\scriptsize I} (21.2 eV) \& He {\scriptsize II} (40.8 eV)
radiations. The resolution of the instrument was set to 450 meV, 5
meV and 10 meV for Al K$\alpha$, He {\scriptsize I} and He
{\scriptsize II} measurements respectively. An open cycle He
cryostat was used to get the temperature variation from room
temperature to 10 K. The base pressure of the chamber was better
than $5\times10^{-11}$ torr during measurements. The sample was
extremely hard with no cleavage plane. We fractured the sample by
putting a top-post to expose clean surface. We also verified the
sample surface by scraping using small grain diamond file. The angle
integrated spectral features obtained from both type of surface
preparations were identical. Reproducibility of the spectra were
ensured after each trial of surface cleaning procedures.

The energy band structure was calculated using full potential
linearized augmented plane wave method within the local density
approximation (LDA) using WIEN2k software.\cite{wien2k} The energy
convergence was achieved using 1000 $k$-points within the first
Brillouin zone. The tetragonal crystal structure\cite{yin} of
HoB$_4$ was considered for the calculations. Convergence was
achieved for the paramagnetic ground state considering both energy
and charge convergence criteria. In order to introduce
electron-electron Coulomb repulsion among Ho 4$f$ electrons, $U$
into the calculation, we have employed LDA+$U$ method considering an
effective electron interaction strength, $U_{eff}$ ($= U - J$; $J$ =
Hund's exchange integral was set to zero following Anisimov et
al.\cite{anisimov}) for various values of $U_{eff}$ upto about 7 eV.

\section{Results and Discussions}

\begin{figure}
 \vspace{-4ex}
\includegraphics [scale=0.4]{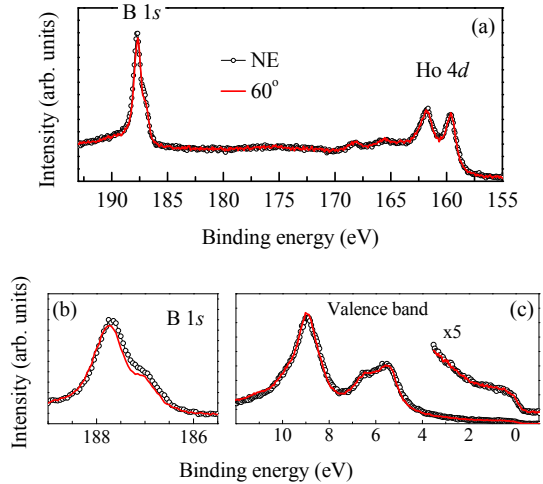}
 \vspace{-30ex}
\caption{(a) B 1$s$ and Ho 4$d$ core level spectra obtained at
normal emission and 60$^o$ angle with surface normal. The expanded
part of B 1$s$ spectra are shown in (b). (c) Valence band spectra
and its rescaled part near $\epsilon_F$ for normal and 60$^o$ angled
emission.}
 \vspace{-4ex}
\end{figure}

Probing depth of the photoemission spectroscopic technique can be
varied by changing the electron emission angle. Following this
procedure, we studied the core level spectra using Al $K\alpha$
photon energy, where the mean escape depth, $\lambda$ for valence
electrons is close to $\sim$~20 \AA.\cite{surface} $\lambda$ can be
made half of its normal emission value if the photoelectron emission
angle is made 60$^o$ with respect to the surface normal. The Ho 4$d$
core level spectra shown in Fig. 2(a) exhibit identical lineshape
for normal emission and 60$^o$ angled emission geometries indicating
similar surface and bulk electronic structures for the electronic
states associated to Ho. B 1$s$ spectra, however, exhibit
modifications with the change in $\lambda$. There is an overall
decrease in intensity of the B 1$s$ peak by about 10\% in the 60$^o$
angled spectrum indicating loss of boron on the surface compared to
bulk. B 1$s$ spectra exhibit two distinct features around 187 and
187.7 eV binding energies - the relative intensity of these features
changes with the change in surface sensitivity suggesting the
surface borons to be somewhat different from the bulk borons.

The valence band spectra shown in Fig. 2(c) are identical at both
the emission geometries. The near Fermi level, $\epsilon_F$ part is
rescaled and shown in Fig. 2(c) exhibiting unaltered lineshape with
the change in surface sensitivity. This is curious as the
intensities near $\epsilon_F$ are largely constituted by B 2$p$
states and surface borons exhibit subtle differences. These results
suggest that the influence of disorder on the conduction electrons
dominates over the translational symmetry breaking occurring at the
surface.

\begin{figure}
 \vspace{-4ex}
\includegraphics [scale=0.4]{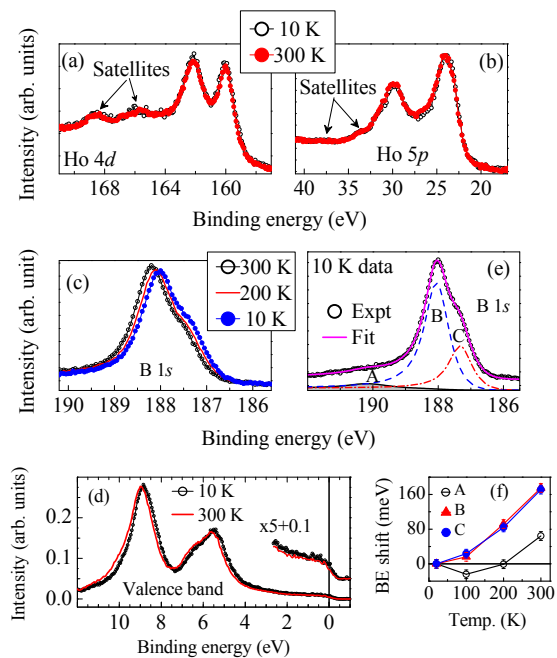}
 \vspace{-16ex}
\caption{Temperature evolution of (a) Ho 4$d$, (b) 5$p$, (c) B 1$s$
and (d) Valence band spectra. (e) Fit to the B 1$s$ spectrum - lines
are the fit curves. (f) Energy shift of the component peaks with
respect to 10 K data.}
 \vspace{-4ex}
\end{figure}

Photoemission being a quantum mechanical process, the spectral
features represent the eigenstates of the excited state Hamiltonian.
If the electron correlation is finite in a system, these final
states, which has a hole due to photoemission, differ significantly
from the initial eigenstates and multiple features appear in the
photoemission spectra associated to different level of screening of
the photoholes, multiplet splitting, etc. The well screened feature
is usually termed as the main peak and other features are called
satellites. Conduction electrons consisting of the hybridized states
of Ho (5$d$6$s$) states and B 2$p$ states are expected to be weakly
correlated as both possess large degree of itineracy due to large
orbital extensions. It is to note here that electron correlation
among Ho 4$f$ electronic states is strong. However, these states
primarily appear far away from $\epsilon_F$ making Ho (5$d$6$s$)
states and B 2$p$ contributions dominant at $\epsilon_F$. Thus,
satellites are not expected in this system. In contrast, the core
level spectra of Ho exhibit significant final state effects with
distinct satellite features. This is found in all the core level
spectra as depicted in Figs. 3(a) and 3(b) for the 4$d$ and 5$p$
spectra, respectively. Presence of the satellites indicates that
electron-electron interaction cannot be neglected in this system as
also observed in various other 4$d$ and 5$d$ systems.\cite{srruo3}
Change in temperature down to 10 K does not influence the spectral
lineshape.

B 1$s$ spectra, on the other hand, are significantly influenced by
the temperature - a gradual shift towards lower binding energy is
observed in Fig. 3(c) with the decrease in temperature, which cannot
be represented by a rigid shift of the experimental spectrum. Three
distinct features, A, B and C are required to simulate the
experimental spectra at any temperature as shown in Fig. 3(e). The
linewidth ($\sim$~0.74 eV), energy separation ($\sim$~0.7 eV) and
relative intensity of the features B and C remain almost the same
across all the temperatures studied. The peak position of the
feature, A exhibit different temperature dependence as shown in Fig.
3(f). The intensity of B relative to C increases in the 60$^o$
angled emission spectrum compared to the normal emission one
suggesting dominant surface contribution in the feature
B.\cite{spatil} Higher binding energy of the surface boron signal
compared to the bulk signal can be associated to surface
reconstruction, defects, surface inhomogeneity that may lead to an
effective reduction of negative boron valency.

There are three inequivalent borons in the lattice as shown in Fig.
1. Since the number of neighbors and bond-lengths are significantly
different for octahedral borons (B1 and B3) and boron dimers (B2),
the Madelung potential is expected to be different for these
non-equivalent sites. Interestingly, the ratio of the integrated
area of the peaks A and C is about 1 : 2.8, which is similar to the
number ratio of B2 and B1+B3 (1 : 3). Thus, the peak A can be
attributed to B2 and C to the photoemission signals from B1 and B3.
B2 dimers bridge the boron octahedra possessing large delocalized
character compared to the octahedral borons leading to a larger
width of the feature A. The binding energy of the B 1$s$ peak
increases with the increase in temperature - the increase is much
less for dimer borons relative to the octahedral ones. Such binding
energy change indicates enhancement of negativity and/or ionicity of
the corresponding borons with the decrease in temperature. Thus,
2$p$ electrons of octahedral borons will gain significant local
character at lower temperatures that could be related to the long
range order in the ground state.

The temperature evolution of the valence band spectra is shown in
Fig. 3(d). The spectral intensities in the vicinity of $\epsilon_F$
appears very similar within the energy resolution of Al $K\alpha$
spectroscopy. The features beyond 5 eV binding energy exhibit a
small but gradual shift towards lower binding energies with the
decrease in temperature (only two temperatures are shown for clarity
in the figure). The energy shift is close to about 160 meV between
300 K and 10 K.

\begin{figure}
 \vspace{-4ex}
\includegraphics [scale=0.4]{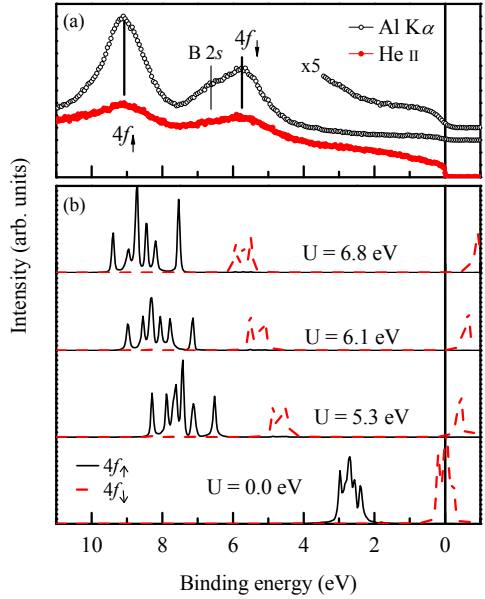}
 \vspace{-16ex}
\caption{(a) Al $K\alpha$ and He {\scriptsize II} valence band
spectra. Near $\epsilon_F$ part of Al $K\alpha$ spectrum is shown in
expanded scale. (b) Calculated partial density of states of Ho 4$f$
up spin (solid line) and down spin (dashed line) are shown for
different values of $U$.}
 \vspace{-4ex}
\end{figure}

Valence band spectra collected using Al $K\alpha$ and He
{\scriptsize II} photon energies are shown in Fig. 4. There are
three distinct features around 5.8, 6.6 and 9.1 eV binding energies.
The intensity of these features enhances significantly at Al
$K\alpha$ photon energy. The comparison of the sensitivity of the
peak intensities with the incident photon energy and the
photoemission cross section\cite{yeh} suggests large Ho 4$f$
photoemission contribution in these peaks. The intensities near
$\epsilon_F$ are primarily due to the B 2$p$ contributions.

In order to clarify these assertions further, the energy band
structure of HoB$_4$ has been calculated within the local spin
density approximations in full potential linearized augmented plane
wave method. Different effective on site Coulomb correlation
strength, $U_{eff}$ for the Ho 4$f$ electrons has been considered to
reproduce the experimental spectra. The calculated results are shown
in Fig. 4(b). It is found that the electron correlation strength of
about 7 eV produces the Ho 4$f_\uparrow$ and 4$f_\downarrow$ partial
density of states at the peak positions observed in the experimental
spectrum.

\begin{figure}
 \vspace{-4ex}
\includegraphics [scale=0.4]{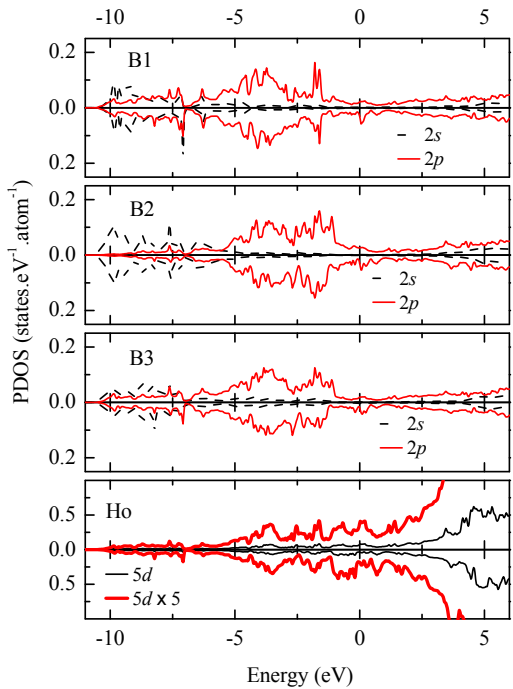}
 \vspace{-10ex}
\caption{B 2$s$, B 2$p$ and Ho 5$d$ partial density of states (PDOS)
from the LSDA calculations. A rescaled Ho 5$d$ PDOS (thick solid
line in the bottom panel) is also shown for clarity.}
 \vspace{-4ex}
\end{figure}

The partial density of states (PDOS) corresponding to B 2$p$ and Ho
5$d$ states from the LDA calculations are shown in Fig. 5. B 2$s$
states dominate between 5 and 10 eV binding energies. Thus, the
features around 5.8 and 9.1 eV in the experimental spectrum, can be
attributed to Ho 4$f_\downarrow$ and 4$f_\uparrow$ contributions,
respectively. The intensities around 6.6 eV is due to B 2$s$
photoemission signal. Boron 2$p$ states are dominated in the energy
range -5 eV to $\epsilon_F$. B1 and B3 2$p$ PDOS have relatively
larger contributions at higher binding energies possessing energy
distributions similar to that of Ho 5$d$ PDOS indicating stronger
hybridization between the octahedral borons with Ho relative to the
coupling of Ho with dimer borons, B2. Interestingly, the B1
contribution at higher binding energies is the largest suggesting
strongest coupling with Ho moments.

\begin{figure}
 \vspace{-4ex}
\includegraphics [scale=0.4]{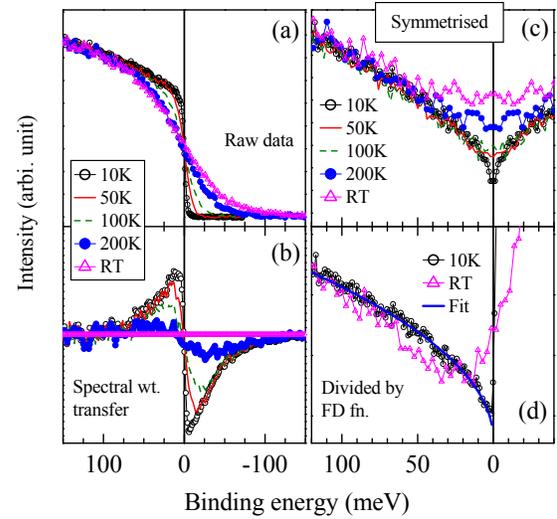}
 \vspace{-24ex}
\caption{(a) High resolution data close to the Fermi level at
different temperatures. (b) Spectral weight transfer calculated by
subtracting the 300 K data from all the raw data. (c) Symmetrized
spectral density of states. (d) Spectral density of states obtained
by the division with the resolution broadened Fermi function.}
 \vspace{-4ex}
\end{figure}

The changes close to $\epsilon_F$ has been investigated with high
energy resolution using He {\scriptsize I} radiations. Since the
electron escape depth at this energy is similar to the 60$^o$ angled
emission in Al K$\alpha$ spectra,\cite{surface} the identical
spectral shape for normal emission and 60$^o$ angled emission shown
in Fig. 2(c) established that the He {\scriptsize I} spectra would
be representative of the bulk electronic structure. The raw data are
shown in Fig. 6(a) exhibiting spectral evolution with significant
spectral weight transfer across $\epsilon_F$ with temperature as
expected in a system of Fermions. Interestingly, the difference
spectra ($I(T) - I(300 K)$) exhibit larger change in spectral weight
above $\epsilon_F$ relative to that below $\epsilon_F$ indicating a
signature of particle-hole asymmetry.\cite{ruth_epl} Intensity at
the Fermi level obtained by the symmetrization of the experimental
spectra ($I(\epsilon - \epsilon_F) + I(\epsilon_F - \epsilon)$)
exhibits gradual decrease with the decrease in temperature. An
additional sharp dip is observed in the 10 K data reflecting
signature of the long range order in the ground state.\cite{dip}
Spectral density of states (SDOS) obtained by the division with the
Fermi-Dirac distribution function exhibit similar lineshape with
$|\epsilon - \epsilon_F|^{0.5}$ dependence at all the temperatures.

Disorder due to the charge degrees of freedom has extensively been
studied theoretically and experimentally.\cite{tvr} Altshuler and
Aronov showed that charge disorder induced effect in a correlated
system exhibit $|\epsilon - \epsilon_F|^{0.5}$ dependence of the
density of states near Fermi level.\cite{altshuler} Such disorder
effect has been observed in varieties of systems such as
Ge$_{1-x}$Au$_x$,\cite{GeAu} Ni(S,Se)$_2$,\cite{dd_nis2}
oxides,\cite{dd_sr2femoo6} {\em etc}. Magnetic fluctuations often
leads to a $|\epsilon - \epsilon_F|^{1.5}$ dependence of the
spectral function\cite{kbm} or $|\epsilon - \epsilon_F|^{0.25}$ in
the vicinity of quantum critical behavior.\cite{VCr} Evidently, the
lineshape of the density of states close to the Fermi level is
sensitive to the electronic interaction parameters as well as
disorder induced effects. In the present case, although the disorder
is with respect to the spin-spin interactions, the spectral function
exhibit a 0.5 exponent, similar to that predicted in
Altshuler-Aronov theory. HoB$_4$ exhibits multiple long range
antiferromganetic order with non-linear magnetization direction. A
change in the temperature and magnetic field influences the
transport and magnetism in this system significantly suggesting
intimate relation between these two phenomena.\cite{song} In
addition, geometrical quadrupolar frustration has been observed in
this compound indicating a strong spin-lattice interaction that
might be the reason for $|\epsilon - \epsilon_F|^{0.5}$ dependence
of the spectral lineshape. It would be interesting to see how the
spectral function evolves in the proximity of spin liquid phase.

\section{Conclusion}

In summary, the electronic structure of a non-collinear
Shastry-Sutherland lattice system, HoB$_4$ is studied using high
resolution photoemission spectroscopy. The results show subtle
differences and loss of boron at the surface compared to the bulk.
Ho core level spectra exhibit signature of satellites suggesting
role of electron correlation among the conduction electrons. The
valence band spectra corresponding to the surface and bulk
electronic structures are found to be similar. The spectral
intensity near the Fermi level possess dominant 2$p$ orbital
character and exhibit disorder induced depletion of density of
states. An additional sharp dip at 10 K is observed that can be
attributed to the long range order in the ground state. Spectral
weight transfer with temperature is found to be asymmetric with
respect to $\epsilon_F$ revealing signature of particle-hole
asymmetry. Spectral lineshape in the vicinity of the Fermi level in
this spin-disordered system exhibit 0.5 exponent, which is similar
to that found in charge disordered system.

\section{Acknowledgements}

The authors N. S. and K. M. acknowledge financial support from the
Dept. of Science and Technology, Govt. of India under the
Swarnajayanti fellowship programme. G. B. acknowledges financial
support from EPSRC, UK (EP/I007210/1).

\end{document}